\documentclass[conference]{IEEEtran}
\IEEEoverridecommandlockouts
\usepackage{cite}
\usepackage{amsmath,amssymb,amsfonts}
\usepackage{algorithmic}
\usepackage{graphicx}
\usepackage{textcomp}
\usepackage{xcolor}
\usepackage{lipsum}
\usepackage{colortbl} 
\usepackage{multirow} 
\usepackage{rotating}
\usepackage{booktabs}
\usepackage{pifont}
\usepackage{hyperref} 
\hypersetup{colorlinks=true, allcolors=blue}
\usepackage{adjustbox}       
\usepackage{array}           
\usepackage{makecell}
\usepackage[utf8]{inputenc}
\usepackage{enumitem}
\usepackage{tcolorbox}
\usepackage{tikz}
\usepackage[normalem]{ulem}
\usepackage{circledsteps}
\usepackage{listings}
\usepackage{pifont}

\newcommand{\etal}{{\em et. al.}}

\newcommand*\circled[2][gray]{\tikz[baseline=(char.base)]{
    \node[shape=circle, draw=#1, fill=#1!20, inner sep=1pt] (char) {#2};}}

\def\BibTeX{{\rm B\kern-.05em{\sc i\kern-.025em b}\kern-.08em
    T\kern-.1667em\lower.7ex\hbox{E}\kern-.125emX}}
\begin{document}

\title{TrojanWhisper: Evaluating Pre-trained LLMs to Detect and Localize Hardware Trojans \vspace{-2mm}\\
}


\author{
    \IEEEauthorblockN{Md Omar Faruque\IEEEauthorrefmark{1}, Peter Jamieson\IEEEauthorrefmark{2}, Ahmad Patooghy\IEEEauthorrefmark{3}, and Abdel-Hameed A. Badawy\IEEEauthorrefmark{1}}
    \IEEEauthorblockA{\IEEEauthorrefmark{1}Klipsch School of ECE, New Mexico State University, Las Cruces, NM 88003, USA}
    \IEEEauthorblockA{\IEEEauthorrefmark{2}Department of Electrical and Computer Engineering, Miami University, Oxford, OH, USA}
    \IEEEauthorblockA{\IEEEauthorrefmark{3}Computer Systems Technology, North Carolina A\&T State University, Greensboro, NC, USA}
    \IEEEauthorblockA{\textit{Emails:} \IEEEauthorrefmark{1}\{faruque, badawy\}@nmsu.edu, \IEEEauthorrefmark{2}jamiespa@miamioh.edu, \IEEEauthorrefmark{3}apatooghy@ncat.edu}
}

\maketitle
\begin{abstract}

Existing Hardware Trojans (HT) detection methods face several critical limitations: logic testing struggles with scalability and coverage for large designs, side-channel analysis requires golden reference chips, and formal verification methods suffer from state-space explosion. The emergence of Large Language Models (LLMs) offers a promising new direction for HT detection by leveraging their natural language understanding and reasoning capabilities. For the first time, this paper explores the potential of general-purpose LLMs in detecting various HTs inserted in Register Transfer Level (RTL) designs, including SRAM, AES, and UART modules. We propose a novel tool for this goal that systematically assesses state-of-the-art LLMs (GPT-4o, Gemini 1.5 pro, and Llama 3.1) in detecting HTs without prior fine-tuning. To address potential training data bias, the tool implements perturbation techniques, \textit{i.e.}, variable name obfuscation, and design restructuring, that make the cases more sophisticated for the used LLMs. Our experimental evaluation demonstrates perfect detection rates by GPT-4o and Gemini 1.5 pro in baseline scenarios (100\%/100\% precision/recall), with both models achieving better trigger line coverage (TLC: 0.82-0.98) than payload line coverage (PLC: 0.32-0.46). Under code perturbation, while Gemini 1.5 pro maintains perfect detection performance (100\%/100\%), GPT-4o (100\%/85.7\%) and Llama 3.1 (66.7\%/85.7\%) show some degradation in detection rates, and all models experience decreased accuracy in localizing both triggers and payloads. This paper validates the potential of LLM approaches for hardware security applications, highlighting areas for future improvement. TrojanWhisper's HT detection results will be available here \href{https://github.com/HSTRG1/TrojanWhisper.git}{https://github.com/HSTRG1/TrojanWhisper.git}.


\end{abstract}


\section{Introduction}
\label{sec:intro}


The globalization of the semiconductor supply chain has created significant security vulnerabilities through the potential insertion of Hardware Trojans (HTs) - malicious modifications inserted by rouge threat actors or rouge Computer-aided design (CAD) tools to integrated circuits (ICs) that can leak information, alter functionality, degrade performance, or deny service~\cite{shakya2017benchmarking}. This security challenge is particularly acute for System-on-Chip (SoC) designers who, faced with time-to-market pressures and resource constraints, increasingly rely on third-party Intellectual Property (3PIP) cores and outsourced design services. These 3PIPs come in three distinct forms: Soft IPs (RTL code in Verilog/VHDL), Firm IPs (synthesized netlists and placed RTL blocks), and Hard IPs (GDSII files and physical layouts).

Early detection of HTs in Soft RTL IPs is crucial, as remediation becomes exponentially costlier in later design stages. This issue is challenging in commercial IP cores, where identifying malicious modifications within thousands of lines of RTL code is formidable~\cite{yasaei2021gnn4tj}. Manual inspection is inadequate for large-scale designs, as RTL HTs blend with legitimate functionality and can evade conventional verification through carefully crafted activation conditions. The taxonomy of Shakya~\etal~\cite{shakya2017benchmarking} highlights the spectrum of RTL HTs, from simple to highly sophisticated mechanisms with extremely low activation probabilities (6.4271e-23), emphasizing the need for automated and scalable detection methods.

 HT detection methods before this can be categorized into four main approaches. Test pattern generation methods~\cite{saha2015improved} attempt to expose HTs through test vectors but cannot guarantee the detection of stealthily triggered HTs. Formal verification approaches~\cite{subramanyan2014formal,rajendran2016formal} convert designs into proof-checking formats but suffer from state explosion and limited detection scope due to predefined properties. Code analysis techniques like FANCI~\cite{waksman2013fanci} and VeriTrust~\cite{zhang2013veritrust} examine RTL code using coverage metrics but require manual analysis and can be bypassed by sophisticated HTs, as DeTrust~\cite{zhang2014detrust} demonstrated. Machine learning (ML) and graph matching approaches have shown promise, with methods ranging from graph similarity~\cite{fyrbiak2019graph} and control-flow matching~\cite{piccolboni2017efficient} to neural networks~\cite{hasegawa2017hardware} and gradient boosting~\cite{han2019hardware}. However, these approaches typically require golden references, extensive feature engineering, or suffer from limited generalization to new RTL designs and HT types. Graph Neural Network (GNN) approaches like GNN4TJ~\cite{yasaei2021gnn4tj} have improved detection capabilities but still face scalability and novel HT detection challenges. 

The limitations of existing approaches motivate our investigation of Large Language Models (LLMs) for more robust and automated HT detection. Given the scarcity of labeled HT datasets, their ability to reason about hardware security implications without extensive domain-specific training data or golden reference designs is crucial~\cite{saha2024llm}. 

In this paper, we present \textbf{\emph{TrojanWhisper}}, a novel LLM approach for HT detection in RTL designs (to the line-level granularity) that leverages state-of-the-art LLMs (GPT-4o, Gemini 1.5 pro, and Llama 3.1) through carefully crafted prompts encoding hardware security domain knowledge. 
According to our assessments, LLMs offer unique advantages in HT detection: 1) They can process RTL code in its native form; 2) Understand structural and semantic properties without explicit feature engineering.  
\underline{Our key contributions include: }
\circled[black]{1} To the best of our knowledge, it is one of the first systematic evaluations of LLMs for HT detection; 
\circled[black]{2} A novel LLM HT signature generation engine for HT detection;
\circled[black]{3} Comprehensive evaluation metrics for assessing LLM HT detection; 
\circled[black]{4} A novel perturbation engine that validates detection robustness through code transformations while maintaining functional equivalence and synthesizability;  
\circled[black]{5} Empirical comparison with existing detection methods.
\definecolor{verilogblue}{RGB}{0,0,255}
\definecolor{verilogpink}{RGB}{128,0,128}
\definecolor{jsonkey}{RGB}{0,0,255}
\definecolor{jsonstring}{RGB}{128,0,0}
\definecolor{jsoncomment}{RGB}{0,128,0}
\definecolor{codebg}{RGB}{236, 236, 236}
\definecolor{highlightbg}{RGB}{255, 255, 150}

\lstdefinestyle{verilogstyle}{
    language=Verilog,
    basicstyle=\ttfamily\scriptsize,
    keywordstyle={\color{verilogblue}},
    breaklines=true,
    showstringspaces=false,
    columns=fixed,
    commentstyle=\color{verilogpink},
    morekeywords=[1]{always, begin, end, if, else, posedge, negedge},
    sensitive=true,
    backgroundcolor=\color{codebg}
}

\lstdefinestyle{jsonstyle}{
    basicstyle=\ttfamily\tiny,
    keywordstyle=\color{jsonkey},
    stringstyle=\color{jsonstring},
    commentstyle=\color{jsoncomment},
    breaklines=true,
    showstringspaces=false,
    numbers=none,
    backgroundcolor=\color{codebg}
}

\section{Background on RTL HT}

HTs at the RTL level represent malicious modifications to the behavioral hardware description that define a circuit's operation and data flow~\cite{shakya2017benchmarking}. Unlike gate- or layout-level HTs that manipulate physical circuit elements, RTL HTs operate at a higher abstraction level by altering the behavioral specification of the design using Hardware Description Languages (HDL) like Verilog or VHDL.

As defined by Shakya~\etal~\cite{shakya2017benchmarking}, these modifications manifest themselves in two fundamental components: a trigger mechanism and a payload. The trigger can be implemented through conditional statements that activate under specific circumstances (\textit{e.g.}, when certain internal states or input patterns occur). These time-based counters activate after a predetermined number of cycles or always-on conditions. The payload, embedded within the RTL code, can manipulate control signals, modify data paths, or introduce unintended state transitions, potentially leading to three primary malicious effects identified in their comprehensive HT taxonomy: 



\circled[black]{1} \underline{\textbf{Functionality Change HT (Type 1):}} 
Once triggered, these HTs modify the circuit's intended behavior by altering or corrupting outputs or computations.
\begin{lstlisting}[style=verilogstyle,
                   caption={Functionality Change HT (UART-T200) in UART receiver.},
                   label={lst:ht1_uart_gemini},
                   numbers=left,
                   numberstyle=\tiny,
                   captionpos=b,
                   frame=none,
                   xleftmargin=2em]
always @(posedge clk) begin 
    if(!resetn) 
        pc <= 8'b0;
    else if (received_data == 8'hAB) 
        pc[0] <= 1'b1;
        
always @(posedge clk) 
    if (fsm_state == FSM_STOP) 
        if (pc[0]) 
            uart_rx_data <= received_data ^ 8'hFF;  
        else 
            uart_rx_data <= received_data;
            
\end{lstlisting}
In this Functionality change HT code snippet~\cite{tjbench} in Listing~\ref{lst:ht1_uart_gemini}, the HT triggers when it detects the specific byte pattern 0xAB (lines 3-9). Once triggered, the payload inverts all subsequently received data until the module is rested (lines 12-13), thus corrupting the UART receiver's functionality.

\circled[black]{2} \underline{\textbf{Information Leakage HT (Type 2):}} These HTs create covert channels to leak sensitive data. An example case of Information Leakage HT~\cite{tjbench} is shown in Fig.~\ref{fig:case_study} (b), the HT activates when address `ADDR'  (0xAA) is read repeatedly (lines 2-8). The payload leaks memory contents by returning data from sequential memory locations (indexed by `mn') instead of the requested address (lines 11-15), creating a covert channel to leak the SRAM contents.

\circled[black]{3} \underline{\textbf{Denial of Service HT (Type 3):}} These HTs aim to degrade or disable hardware functionality, making it partially or entirely unusable.

\begin{lstlisting}[style=verilogstyle,
                   caption={Denial of Service HT (AES-T220) in AES-128 encryption.},
                   label={lst:ht3_aes},
                   numbers=left,
                   numberstyle=\tiny,
                   captionpos=b,
                   frame=none,
                   xleftmargin=2em]
always @(posedge clk) 
    if (reset) 
        pq <= 8'h00; 
    else if (state[7:0] == 8'hAA && key[7:0] == 8'h55) 
        pq <= pq + 1;
    else 
        pq <= 8'h00;
        
always @ (posedge clk) 
    if (reset) begin
        s0 <= 128'h0;
        k0 <= 128'h0;
    end  else if (pq != 8'hFF) begin
        s0 <= state ^ key;
        k0 <= key;
    end
\end{lstlisting}

As shown in the code snippet in Listing~\ref{lst:ht3_aes}, the Denial of Service HT (example taken from~\cite{tjbench}) triggers when it detects a specific pattern in the input `state' (plaintext) and `key' (lines 3-9). When the counter `pq' reaches 0xFF, the payload blocks the initial XOR operation of the AES encryption (lines 11-18), effectively denying the encryption service.


\section{Proposed Methodology}

\label{sec:methodology}
Our research methodology consists of three main components: 
\circled[teal]{A} \textbf{HT Signature Generation Engine}, 
\circled[teal]{B} \textbf{Perturbation Generation Engine},  and \circled[teal]{C} \textbf{LLM HT Detection Engine}, which will be discussed in the following subsection.


\subsection{HT Signature Generation Engine}

We propose an iterative HT signature generation and ranking methodology that systematically learns detection patterns from the Trust-Hub~\cite{trusthub_chiptrojan} dataset $D = \{(C_i, T_i, M_i)\}$, comprising clean RTL code $C_i$, HT-infected code $T_i$, and metadata $M_i$. Our multi-stage pipeline employs an LLM through function $L(C_i, T_i, M_i)$ to extract HT signatures $P = \{p_1, p_2, ..., p_m\}$ for both trigger mechanisms and payload patterns. These signatures undergo refinement $R(P) \rightarrow S'$ using similarity metric $\delta(p_x, p_y) > \theta$, where similar signatures are merged and generalized to improve cross-variant detection.

Each signature $s'_j$ undergoes validation through $V(s'_j)$ against a separate validation set $D_{val}$, producing performance vectors $v_j = [\alpha_j, \beta_j, \gamma_j]$ that capture detection rate, false positive rate, and generalization capability. A ranking system $W(v_j)$ assigns weights $w_j$ based on these metrics, creating an ordered set $S = \{(s_1, w_1), ..., (s_q, w_q)\}$. The system implements continuous improvement through an iterative process $I(S, F)$, analyzing failure cases to refine existing signatures.

To address emerging threats, our system incorporates a zero-day detection mechanism $N(Z)$ that analyzes novel HT patterns $Z$ and generates corresponding signatures $s_{new}$, where $S_{t+1} = S_t \cup \{s_{new}\}$. This creates a dynamic, hierarchical signature database that continuously adapts to new threats while maintaining high detection reliability through systematic validation and ranking.


\subsection{Perturbation Generation Engine}

Our perturbation engine powered by an LLM agent addresses a challenge in LLM HT detection: the potential overlap between open-source HT RTL designs and the models' training data. This overlap could lead to unreliable detection results based on memorization rather than understanding hardware security principles. The engine implements a systematic approach to code transformation while maintaining functional equivalence and synthesizability. At its core, the framework applies \textbf{three key transformation strategies}. 

\noindent\circled[black]{1} \underline{Variable name obfuscation:} First, it performs comprehensive variable name obfuscation, replacing meaningful identifiers with arbitrary combinations of letters while preserving module interfaces and port names. This tests the LLMs' understanding of circuit functionality independent of descriptive naming conventions. 

\noindent\circled[black]{2} \underline{Addition of synthesizable redundant logic:}  It then introduces additional synthesizable logic structures, including redundant registers and alternative combinational implementations, which challenge the models to maintain accurate detection despite these variations. 

\noindent\circled[black]{3} \underline{Restructuring:} Lastly, it restructures state machines and control logic while preserving original functionality, testing the models' comprehension of behavioral equivalence. Critically, all transformations adhere to strict synthesizability constraints, ensuring modified designs remain implementable in hardware.

\subsection{LLM HT Detection Engine }


The LLM detection engine combines HT signature-based identification with in-context learning for HT analysis. The engine first applies the detected HT signatures derived from our HT Signature Generation Engine to identify potential trigger mechanisms and payload patterns in the RTL code. These detection HT signatures, generated through systematic analysis of HT characteristics, help pinpoint suspicious circuit elements and behavioral patterns. The engine employs an in-context learning approach for HT classification by providing type-specific examples and attributes for each HT category (functionality modification, information leakage, and denial of service). The system processes RTL designs through this two-stage analysis: first applying detection HT signatures to identify the presence of HTs and locate suspicious components, then using contextual understanding to classify the type of identified HTs. Results are structured as XML outputs detailing the number of potential detected HTs, their identified trigger/payload components, and HT categorization.








\section{Evaluation Metrics}
\label{sec:matrices}

Our evaluation framework employs a set of metrics to assess three aspects of HT detection in RTL IP cores: \circled[black]{1} \underline{core detection capabilities}, \circled[black]{2} \underline{line localization precision}, and \circled[black]{3} \underline{HT category classification accuracy}. These metrics provide a detailed, multi-dimensional evaluation of agent performance across various levels. Our test dataset consists exclusively of HT-infected RTL modules with no HT-free designs. However, our line-level approach allows for an effective evaluation, as most of the codebase represents clean design code, creating a ``negative'' class for detection. The following subsections describe these metrics.



\subsection{HT Detection Metrics}

 HT detection metrics are based on classification methods to evaluate the agent’s ability to detect HTs at the RTL IP core level (Verilog file).  Therefore, we call these broad metrics to determine whether an agent detects hidden HTs. We will describe the True Positives, False Positives, and False Negatives for these metrics from the perspective of how good the agent is classifying the existence of an HT where the number of HTs in a design is set to the parameter $k$. 

\noindent\textbf{1) True Positives (TP)}: Count of \textbf{successfully detected} (either trigger or payload signatures Fig. ~\ref{fig:case_study} (a)) HTs. For example, if a module contains 4 HTs ($k = 4$) and the agent detects all 4, \( TP = 4 \). TP requires correct identification of triggers (such as specific signals or states for activation) or payloads (like malicious modifications to signals or information flows).

\noindent\textbf{2) False Positives (FP)}: Instances where clean RTL code is mistakenly flagged as containing HTs. For example, if the agent claims to find 6 HTs in code with a $k = 4$ in the actual design, \( FP = 2 \) (indicating two false alarms). Lowering FP reduces unnecessary investigations, enhancing efficiency.

\noindent\textbf{3) False Negatives (FN)}: HTs that are present in the RTL but missed by the agent. If there are 4 HTs and only two are identified, \( FN = 2 \), indicating missed detections that can lead to security risks.



We provide these factors as a tuple \{$k$, \#TP, \#FP, \#FN\} such that the closer the 2nd number is to $k$ and the lower \#FP and \#FN suggest the tool is better.  
Note that there is no true negative classification at this granularity level since the benchmarks entirely consist of HT-infected RTL designs. 

\subsection{Localization Precision Metrics}

Localization precision provides a line-by-line analysis of the agent's ability to classify Verilog RTL Lines-of-Code (LoC).  These metrics describe how accurately the agent identifies specific HT lines (which include a distinction between the payload and trigger) in the code.  In this regard, we note that when compared to the structural netlist implementations, RTL behavioral designs do not have a simple one-to-one relationship.  A single line of RTL code can correspond to many nodes within the structural netlist.  Similarly, inferred structures such as Finite State Machines (FSMs) and inferred memories in synthesizable Verilog might result in non-obvious LoC mappings.  Because of these, this classification approach describes an agent's capability of detection but is pinpoint accurate as an identification of HTs.

Our confusion matrix for this purpose is defined by TRUENESS reflecting the correct classification and POSITIVENESS indicating whether a line has the presence of an HT or not.

\begin{itemize}
    \item $LoC_{TP_{payload}}$: This parameter is a numerical count of the number of lines identified correctly as the payload of an HT.
    \item $LoC_{TP_{trigger}}$: This parameter is a numerical count of the number of lines identified correctly as the trigger of an HT.
    \item $LoC_{TP_{clean}}$: This parameter is a numerical count of the number of lines identified correctly as clean RTL code.
    \item $LoC$: This parameter is a numerical count of the number of lines of RTL code in the design.
\end{itemize}

\begin{itemize}
    \item \textbf{Trigger Line Coverage (TLC)}: Assesses how accurately the agent identifies the precise RTL lines of code related to the HT trigger. TLC is a percentage calculated by
   \begin{equation}
       TLC = \frac{LoC_{TP_{trigger}}}{LoC_{TP_{trigger}} + LoC_{FP_{trigger}}}
   \end{equation} 

    \item \textbf{Payload Line Coverage (PLC)}: Measures the agent’s accuracy in identifying specific LoC related to HT payloads, calculated similarly to TLC.
    \begin{equation}
       PLC = \frac{LoC_{TP_{payload}}}{LoC_{TP_{payload}} + LoC_{FP_{payload}}}
   \end{equation} 

    \item \textbf{Accuracy of Classification (AC)} is a measure of how accurate the agent is and is calculated as
\begin{equation}
    AC = \frac{LoC_{TP_{payload}} + LoC_{TP_{trigger}} + LoC_{TP_{clean}}}{LoC}
\end{equation}
\end{itemize}

The final metric, AC, provides a summary value, allowing us to compare the quality of LLM detection between systems.   


\subsection{HT Category Classification Accuracy Metric}

Our last metric evaluates how effective the agent is at providing a deeper understanding of what the HT is doing based on the known taxonomy of HTs.

\textbf{HT Category Classification Accuracy (TCCA)} measures the LLM's ability to classify an HT correctly. Taking  Type-2 (Information Leakage) HTs as an example, TCCA evaluates both correctly classified Type-2 HTs and false positives where non-Type-2 HTs or legitimate components are incorrectly classified as Type-2. In a design with six  Type-2 HTs ($k=6$), if the LLM detects and correctly classifies 4 Type-2 HTs but also incorrectly flags 2  (either other HT types or legitimate circuits) as Type-2, the TCCA\_Type2 would be 4/(4+2) = 66.7\%.

\begin{figure*}[t]
\centering
\includegraphics[width=1.0\textwidth]{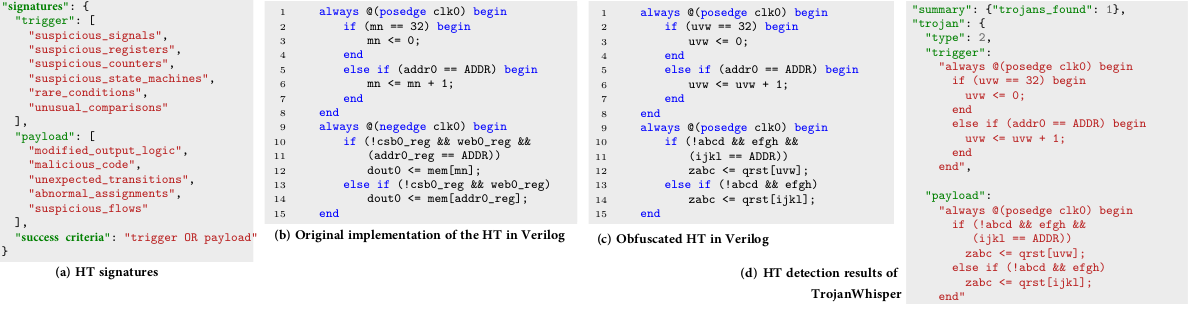}
\vspace{-4mm}
\caption{TrojanWhisper detection of an information leakage HT in SRAM design (SRAM-T110~\cite{tjbench}): HT signatures (a), original HT (SRAM-T110) implementation  (b), obfuscated
HT (c), and detection results (d).}
\label{fig:case_study}
\vspace{-4mm}
\end{figure*}

\section{Experimental Results}
\label{sec:results}

We use Icarus Verilog (version 11.0) to ensure functional equivalence of the obfuscation designs with the original implementation, confirming that our obfuscation techniques preserve the intended circuit behavior. 
Our main system components are built in a Python 3.10.14 environment with the core LLM interactions handled through multiple API interfaces: OpenAI's GPT-4o API, Google's Generative AI SDK (version 0.8.3) for Gemini 1.5 Pro, and the Groq Client library for LLama3.1. Environment management uses python-dotenv (1.0.1) for secure API key handling, while XML processing relies on the ElementTree library from Python's standard library.

Our framework leverages multiple LLM backends. Table~\ref{tab:llm_config} presents the detailed configuration parameters for each model.
Our experimental dataset~\cite{tjbench} consists of 14 HT benchmarks, which have been derived from three clean hardware designs sourced from GitHub repositories: an SRAM controller (52 lines), UART communication module (430 lines), and AES-128 cryptographic core (768 lines). These clean designs were modified to incorporate HTs, creating 14 distinct HT-infected variants. These HT-inserted designs were uploaded to GitHub after the knowledge cutoff date of the evaluated LLM agents, ensuring they were not part of the model's training data. As shown in Table~\ref{tab:ht_categorization}, the dataset comprises a distribution of three different HT types, with 4 Type-1, 4 Type-2, and 6 Type-3 HTs.

We used these 14 HT-infected designs as our baseline dataset to create our test cases. Additionally, we applied our perturbation engine to each of these 14 designs to generate corresponding obfuscated versions while maintaining functional equivalence, resulting in a complementary dataset for evaluating detection robustness under code transformations.

\begin{table}[t]
\renewcommand{\arraystretch}{1}
\caption{Hardware Trojan Benchmark Distribution}
\label{tab:ht_categorization}
\setlength{\tabcolsep}{6pt}
\centering
\small
\begin{adjustbox}{max width=0.85\columnwidth}
\begin{tabular}{l|c|l}
\toprule
\rowcolor[gray]{0.85}
\textbf{HT Type} & \textbf{Count} & \textbf{Benchmark IDs} \\
\midrule
Type-1 & \multirow{2}{*}{4} & SRAM-HT100, UART-HT100, \\
& & AES-HT100, AES-HT110 \\
\cmidrule{1-3}
Type-2 & \multirow{2}{*}{4} & SRAM-HT220, AES-HT210, \\
& & AES-HT220, UART-HT200 \\
\cmidrule{1-3}
Type-3 & \multirow{2}{*}{6} & SRAM-HT120, SRAM-HT320, AES-HT120, \\
& & UART-HT110, UART-HT220, SRAM-HT110 \\
\bottomrule
\end{tabular}
\end{adjustbox}
\end{table}

The case study shown in Fig.\ref{fig:case_study} demonstrates the complete workflow of TrojanWhisper through its key components. Fig.\ref{fig:case_study}(a) shows the top HT signatures generated by our HT Signature Generation Engine. These signatures are organized into two categories: trigger patterns (suspicious signals, registers, counters, state machines, rare conditions, and unusual comparisons) and payload patterns (modified output logic, malicious code, unexpected transitions, abnormal assignments, and suspicious flows). When applied to an SRAM information leakage HT from our test benchmark shown in Fig.\ref{fig:case_study}(b), these signatures effectively identify both the counter-based trigger mechanism using the variable ``mn'' and the payload that manipulates memory access patterns. To validate detection robustness, our Perturbation Generation Engine transformed the code as shown in Fig.\ref{fig:case_study}(c), applying variable name obfuscation (\textit{e.g.}, ``mn'' to ``uvw'', ``mem'' to ``qrst''). Fig.\ref{fig:case_study}(d) shows our detection engine's analysis output, which successfully identifies this as a Type-2 information leakage HT by detecting both the suspicious counter-based trigger and the abnormal memory access pattern in the payload, despite the obfuscation attempts.

\begin{table}[t]
\renewcommand{\arraystretch}{1.0}
\vspace{-2mm}
\caption{LLM Configuration Parameters}
\label{tab:llm_config}
\vspace{-2ex}
\setlength{\tabcolsep}{4pt}
\centering
\small
\begin{adjustbox}{max width=0.85\columnwidth}
\begin{tabular}{l|c|c|c}
\toprule
\rowcolor[gray]{0.85}
\textbf{Parameter} & \textbf{GPT-4o} & \textbf{Gemini 1.5} & \textbf{Llama 3.1} \\
& & \textbf{Pro} & \textbf{Versatile} \\
\midrule
Model Version & GPT-4o & Gemini-1.5-pro-latest & llama-3.1-70b-versatile \\
\cmidrule{1-4}
Temperature & 1.0 & 1.0 & 1.0 \\
\cmidrule{1-4}
Top-p & 1.0 & 0.95 & 1.0 \\
\cmidrule{1-4}
Max Input Tokens & 128K & \textasciitilde{}200k & 128k \\
\cmidrule{1-4}
Max Output Tokens & \textasciitilde{}16K  & 8K & \textasciitilde{}32k \\
\cmidrule{1-4}
Knowledge Cutoff & Oct 2023 & Nov 2023 & Dec 2023 \\
\bottomrule
\end{tabular}
\end{adjustbox}
\vspace{-5mm}
\end{table}

\begin{table}[t]
\renewcommand{\arraystretch}{0.9}
\caption{HT Detection Results Without Obfuscation/Perturbation}
\label{tab:no-obfuscation}
\vspace{-1ex}
\setlength{\tabcolsep}{3.5pt}
\centering
\small
\begin{adjustbox}{max width=1\columnwidth}
\begin{tabular}{c|c|c|c|c|c|c|c}
\toprule
\multirow{2}{*}{\textbf{Model}} & \multirow{2}{*}{\textbf{Test Category}} & \multicolumn{1}{c|}{\textbf{Detection Performance}} & \multicolumn{3}{c|}{\textbf{Localization Precision}} & \multicolumn{2}{c}{\textbf{Type Classification}} \\
\cmidrule(lr){3-3} \cmidrule(lr){4-6} \cmidrule(lr){7-8}
\rowcolor[gray]{0.85}
& & \textbf{\{k, TP, FP, FN\}} & \textbf{TLC} & \textbf{PLC} & \textbf{AC} & \textbf{TCCA} & \textbf{Type Analysis} \\
\midrule
\multirow{4}{*}{\raisebox{2em}{\rotatebox[origin=c]{0}{\textbf{GPT-4o}}}} & HT1 Cases & \cellcolor{green!25}\{4, 4, 0, 0\} & 1.00 & 0.29 & 0.96 & 1.000 & 4/4 Type-1 \\
& HT2 Cases & \cellcolor{green!25}\{4, 4, 0, 0\} & 0.94 & 0.62 & 0.89 & 0.750 & 3/4 Type-2 \\
& HT3 Cases & \cellcolor{green!25}\{6, 6, 0, 0\} & 1.000 & 0.19 & 0.90 & 0.500 & 3/6 Type-3 \\[2pt]
\cmidrule{2-8}
\rowcolor{cyan!10}\textit{Aggregate} & & \textbf{\{14, 14, 0, 0\}} & \textbf{0.98} & \textbf{0.32} & \textbf{0.92} & \textbf{0.714} & \textbf{---} \\
\midrule
\multirow{4}{*}{\raisebox{2em}{\rotatebox[origin=c]{0}{\textbf{ 
 Gemini 1.5}}}} & HT1 Cases & \cellcolor{green!25}\{4, 4, 0, 0\} & 1.00 & 0.40 & 0.93 & 0.750 & 3/4 Type-1 \\
& HT2 Cases & \cellcolor{green!25}\{4, 4, 0, 0\} & 0.75 & 1.00 & 0.87 & 0.750 & 3/4 Type-2 \\
& HT3 Cases & \cellcolor{green!25}\{6, 6, 0, 0\} & 0.84 & 0.34 & 0.88 & 0.333 & 2/6 Type-3 \\[2pt]
\cmidrule{2-8}
\rowcolor{cyan!10}\textit{Aggregate} & & \textbf{\{14, 14, 0, 0\}} & \textbf{0.82} & \textbf{0.46} & \textbf{0.89} & \textbf{0.571} & \textbf{---} \\
\midrule
\multirow{4}{*}{\raisebox{2em}{\rotatebox[origin=c]{0}{\textbf{ 
 Llama 3.1}}}} & HT1 Cases & \cellcolor{green!25}\{4, 4, 0, 0\} & 0.57 & 0.23 & 0.90 & 0.750 & 3/4 Type-1 \\
& HT2 Cases & \cellcolor{yellow!25}\{4, 4, 4, 0\} & 0.65 & 0.30 & 0.81 & 0.375 & 3/8 Type-2 \\
& HT3 Cases & \cellcolor{green!25}\{6, 6, 0, 0\} & 1.00 & 0.67 & 0.89 & 0.333 & 2/6 Type-3 \\[2pt]
\cmidrule{2-8}
\rowcolor{cyan!10}\textit{Aggregate} & & \textbf{\{14, 14, 4, 0\}} & \textbf{0.72} & \textbf{0.33} & \textbf{0.87} & \textbf{0.444} & \textbf{---} \\
\bottomrule
\end{tabular}
\end{adjustbox}
\end{table}

\subsection{Performance Evaluation}
\label{subsec:PerfEval}

\begin{table}[t]
\renewcommand{\arraystretch}{0.9}
\caption{HT Detection Results With Obfuscation/Perturbation}
\label{tab:with-obfuscation}
\vspace{-1ex}
\setlength{\tabcolsep}{3.5pt}
\centering
\small
\begin{adjustbox}{max width=1\columnwidth}
\begin{tabular}{c|c|c|c|c|c|c|c}
\toprule
\multirow{2}{*}{\textbf{Model}} & \multirow{2}{*}{\textbf{Test Category}} & \multicolumn{1}{c|}{\textbf{Detection Performance}} & \multicolumn{3}{c|}{\textbf{Localization Precision}} & \multicolumn{2}{c}{\textbf{Type Classification}} \\
\cmidrule(lr){3-3} \cmidrule(lr){4-6} \cmidrule(lr){7-8}
\rowcolor[gray]{0.85}
& & \textbf{\{k, TP, FP, FN\}} & \textbf{TLC} & \textbf{PLC} & \textbf{AC} & \textbf{TCCA} & \textbf{Type Analysis} \\
\midrule
\multirow{4}{*}{\raisebox{2em}{\rotatebox[origin=c]{0}{\textbf{GPT-4o}}}} & HT1 Cases & \cellcolor{green!25}\{4, 4, 0, 0\} & 0.87 & 0.24 & 0.93 & 1.000 & 4/4 Type-1 \\
& HT2 Cases & \cellcolor{yellow!25}\{4, 2, 0, 2\} & 0.71 & 0.33 & 0.81 & 1.000 & 2/2 Type-2 \\
& HT3 Cases & \cellcolor{green!25}\{6, 6, 0, 0\} & 0.90 & 0.14 & 0.90 & 0.500 & 3/6 Type-3 \\[2pt]
\cmidrule{2-8}
\rowcolor{cyan!10}\textit{Aggregate} & & \textbf{\{14, 12, 0, 2\}} & \textbf{0.86} & \textbf{0.20} & \textbf{0.88} & \textbf{0.750} & \textbf{---} \\
\rowcolor{cyan!10} & & (-2 TP) & ↓ & ↓ & ↓ & ↑ & \\
\midrule
\multirow{4}{*}{\raisebox{2em}{\rotatebox[origin=c]{0}{\textbf{Gemini 1.5}}}} & HT1 Cases & \cellcolor{green!25}\{4, 4, 0, 0\} & 0.69 & 0.29 & 0.93 & 0.750 & 3/4 Type-1 \\
& HT2 Cases & \cellcolor{green!25}\{4, 4, 0, 0\} & 0.78 & 0.89 & 0.87 & 0.750 & 3/4 Type-2 \\
& HT3 Cases & \cellcolor{green!25}\{6, 6, 0, 0\} & 0.73 & 0.23 & 0.86 & 0.667 & 4/6 Type-3 \\[2pt]
\cmidrule{2-8}
\rowcolor{cyan!10}\textit{Aggregate} & & \textbf{\{14, 14, 0, 0\}} & \textbf{0.73} & \textbf{0.33} & \textbf{0.89} & \textbf{0.643} & \textbf{---} \\
\rowcolor{cyan!10} & & (→) & ↓ & ↓ & → & ↑ & \\
\midrule
\multirow{4}{*}{\raisebox{2em}{\rotatebox[origin=c]{0}{\textbf{Llama 3.1}}}} & HT1 Cases & \cellcolor{yellow!25}\{4, 4, 2, 0\} & 0.67 & 0.17 & 0.91 & 0.333 & 2/6 Type-1 \\
& HT2 Cases & \cellcolor{red!25}\{4, 2, 1, 2\} & 0.67 & 0.33 & 0.81 & 0.333 & 1/3 Type-2 \\
& HT3 Cases & \cellcolor{yellow!25}\{6, 6, 3, 0\} & 0.79 & 0.33 & 0.88 & 0.111 & 1/9 Type-3 \\[2pt]
\cmidrule{2-8}
\rowcolor{cyan!10}\textit{Aggregate} & & \textbf{\{14, 12, 6, 2\}} & \textbf{0.72} & \textbf{0.25} & \textbf{0.87} & \textbf{0.222} & \textbf{---} \\
\rowcolor{cyan!10} & & (+2 FP, +2 FN) & → & ↓ & → & ↓ & \\
\bottomrule
\end{tabular}
\vspace{-3mm}
\end{adjustbox}
\footnotesize{\textbf{Legend:} ↑ Increase from baseline, ↓ Decrease from baseline, → No significant change}
\vspace{-5mm}
\end{table}

Our experimental results are presented in two tables: Table~\ref{tab:no-obfuscation} shows detection performance without obfuscation/perturbation, and Table~\ref{tab:with-obfuscation} presents results with these techniques applied. 
We first calculated the individual row metrics (HT1, HT2, HT3 Cases) using only the samples of that particular HT type - for instance, HT1 cases consider only the 4 Type-1 HTs, with their detection performance tuple $\{k=4, \text{TP}, \text{FP}, \text{FN}\}$ and corresponding TLC, PLC, AC, and TCCA metrics computed for just these samples. We then considered all HT samples together rather than averaging across HT types for the aggregate metrics. For the detection performance tuple {k, TP, FP, FN}, we summed the total number of HTs (k=14) across all samples, along with the total number of true positives (TP), false positives (FP), and false negatives (FN) identified across all samples. Similarly, for metrics like TLC, PLC, AC, and TCCA, we considered the total correctly identified lines or HT-type classifications across all 14 samples divided by the total number of relevant lines or HT-type classifications, rather than taking the average of the three category-specific metrics.


\noindent\circled[black]{1} \underline{\textbf{Baseline Case:}} 
In the baseline evaluation shown in Table~\ref{tab:no-obfuscation}, GPT-4o and Gemini 1.5 demonstrate perfect detection rates, successfully identifying all 14 HTs without any FPs or FNs in their aggregate performance. Llama 3.1, while detecting all true HTs, generated four FPs in HT2 test cases, resulting in a performance tuple of $\{14, 14, 4, 0\}$. For localization precision, GPT-4o achieved the highest trigger line coverage (TLC) of 0.98, though its payload line coverage (PLP) was lower at 0.32. Gemini 1.5 showed more balanced but slightly lower localization metrics with TLC of 0.82 and PLP of 0.46. The AC is strong across all models, with GPT-4o achieving 0.92, Gemini 1.5 at 0.89, and Llama 3.1 maintaining 0.87, indicating robust performance in correctly identifying clean versus HT-infected code. However, the HT classification accuracy (TCCA) varied significantly across models, with GPT-4o achieving 0.714, followed by Gemini 1.5 at 0.571, and Llama 3.1 at 0.444, suggesting that while overall detection and classification accuracy (line) was robust and precise categorization of HT types remains challenging. 

\noindent\circled[black]{2} \underline{\textbf{Perturbed Case:}} 
The introduction of obfuscation techniques (Table~\ref{tab:with-obfuscation}) provides a more reliable assessment of true detection capabilities by mitigating potential training data advantages. Here, we see different performance across models. Gemini 1.5 maintained a perfect detection rate with an ideal performance tuple of $\{14, 14, 0, 0\}$  even under obfuscation. In contrast, Llama 3.1's and GPT-4o's performance showed some degradation, particularly in HT2 cases where both the agents missed two HTs (FN=2). 
All models experienced decreased trigger line coverage (TLC), with GPT-4o dropping to 0.86 from 0.98, Gemini 1.5 to 0.73 from 0.82, and Llama 3.1 remained stable at 0.72. Payload line coverage (PLC) similarly declined across models. Notably, the accuracy of classification (AC) remained stable under perturbation (GPT-4o: 0.88, Gemini 1.5: 0.89, Llama 3.1: 0.87), though Trojan categorization accuracy (TCCA) varied significantly, with Llama 3.1 showing the most degradation (0.222 from 0.444) while GPT-4o slightly improved (0.750 from 0.714).
These results suggest that while code obfuscation impacts the models' ability to precisely locate HTs (as shown by decreased TLC and PLC), their overall ability to distinguish between clean and infected code (AC) remains robust. However, accurate categorization of HT types becomes more challenging for some models.

\begin{table}[t]
\renewcommand{\arraystretch}{1.0}
\caption{Computational Cost Analysis Across LLM Models}
\label{tab:cost_analysis}
\vspace{-2ex}
\setlength{\tabcolsep}{4pt}
\centering
\small
\begin{adjustbox}{max width=0.85\columnwidth}
\begin{tabular}{l|c|c|c}
\toprule
\rowcolor[gray]{0.85}
\textbf{Parameter} & \textbf{GPT-4o} & \textbf{Gemini 1.5 Pro} & \textbf{Llama 3.1 Versatile} \\
\midrule
Avg Time/Sample (s) & 6.01 & 10.91 & 2.63 \\
\cmidrule{1-4}
Input Tokens/Sample & 1437.2 & N/A & 2499.9 \\
\cmidrule{1-4}
Output Tokens/Sample & 1406.36 & N/A & 3985.14 \\
\cmidrule{1-4}
Cost/Sample (\$) & 0.0079 & free-tier & free-tier \\
\bottomrule
\end{tabular}
\end{adjustbox}
\vspace{1mm}
\begin{tabular}{p{0.9\columnwidth}}
\footnotesize{
\textbf{Note:} 
N/A indicates token counts are unavailable. 
}
\end{tabular}
\vspace{-5mm}
\end{table}

Our evaluation measured LLM inference costs across the three LLM agents for the HT detection task using the obfuscated benchmarks, as shown in Table~\ref{tab:cost_analysis}. The results show revealing tradeoffs: Gemini 1.5 Pro achieved perfect detection using the free tier but needed 10.91s per sample. Llama 3.1 was fastest (2.63s) and free but dropped to 66.7\% precision, 85.7\% recall. GPT-4o struck balance - 6.01s and \textdollar0.0079 per sample while maintaining 100\% precision, 85.7\% recall. All times include API latency and execution times for results extraction python scripts.


\noindent\circled[black]{3} \underline{\textbf{Key Takeaways:}} 
These results, shown in Tables~\ref{tab:no-obfuscation} and~\ref{tab:with-obfuscation}, underscore the promise and limitations of LLM approaches for HT detection.

\begin{table}[b]
\renewcommand{\arraystretch}{1.0}
\vspace{-5mm}
\caption{Comparison of HT Detection Methods}
\label{tab:comparison}
\vspace{-2ex}
\setlength{\tabcolsep}{4pt}
\centering
\small
\begin{adjustbox}{max width=0.85\columnwidth}
\begin{tabular}{l|l|c|c|c}
\toprule
\rowcolor[gray]{0.85}
\textbf{Work} & \textbf{Method/} & \textbf{Detection} & \textbf{HT} & \textbf{Time/}\\
\textbf{[Venue, Year]} & \textbf{Technique} & \textbf{Rate (\%)} & \textbf{Local.} & \textbf{Sample}\\
\midrule
Our work & LLM & 100/100 & \checkmark & 10.91s\\
\cmidrule{1-5}
MLCAD'24~\cite{fan2024efficient} & GBDT & 99/98 & \checkmark & 0.045s\\
\cmidrule{1-5}
ITCA'24~\cite{li2024pinpointing} & NLP-Based &  97.94/97.77 & \checkmark & --\\
\cmidrule{1-5}
TVLSI'22~\cite{yasaei2022golden} & GCN & 98.9/88.4 & \checkmark & 10.2ms\\
\cmidrule{1-5}
TCAD'23~\cite{hassan2023circuit} & Topology-GNN & 93.35/91.38 & \ding{55} & 0.38s\\
\cmidrule{1-5}
DATE'21~\cite{yasaei2021gnn4tj} & GNN & 92/97 & \ding{55} & 21.1ms\\
\cmidrule{1-5}
ISCAS'19~\cite{han2019hardware} & Gradient Boost & --/100 & \ding{55} & 1.36s\\
\cmidrule{1-5}
ICCIT'19~\cite{islam2019socio} & Network Analysis & 98/98 & \checkmark & --\\
\cmidrule{1-5}
AsianHOST'18~\cite{zareen2018detecting} & ML-Immune & 87/85 & \ding{55} & --\\
\cmidrule{1-5}
TECS'17~\cite{piccolboni2017efficient} & Graph Match & --/100 & \ding{55} & 5.02s\\
\cmidrule{1-5}
ITC'17~\cite{nahiyan2017hardware} & Flow Analysis & --/100 & \ding{55} & 292.85s\\
\cmidrule{1-5}
HLDVT'17~\cite{demrozi2017exploiting} & ML/Graph Iso & --/100 & \ding{55} & 1.15s\\
\bottomrule
\end{tabular}
\end{adjustbox}
\begin{tabular}{p{0.95\columnwidth}}
\footnotesize{
\textbf{1) Detection Rate:} Shown as Precision/Recall where available; `--' indicates not reported;
\textbf{2) HT Local.:} \checkmark: Supports localization, \ding{55}: No localization support
}
\end{tabular}
\end{table}

\noindent\underline{\textbf{Generalization vs. Memorization:}} 
Gemini 1.5 pro's ability to maintain perfect detection performance even under obfuscation (while other models degraded) demonstrates that LLMs can perform HT detection beyond a simple comparison with the memorized golden reference design.


\noindent\underline{\textbf{Detection-Localization Gap:}}
 While models maintain good line detection capabilities (high AC scores of 0.87-0.89), their ability to precisely locate HTs (TLC/PLC) degrades under obfuscation. This suggests LLMs are better at identifying the presence of threat comparisons than pinpointing their exact location.

\noindent\underline{\textbf{Viability for Security Applications:}} 
 The consistently high detection rates and stable accuracy classification (AC) scores across obfuscated scenarios validate LLM approaches as a promising complement to existing HT detection methods.




\subsection{Comparison with other Detection Methods}

Compared to existing approaches in the literature, as shown in Table~\ref{tab:comparison}, our LLM-based method using Gemini 1.5 pro achieves superior performance with 100/100\% (Precision/Recall) detection rate while providing HT localization capabilities. Among neural network-based approaches, GNN variants show strong but varying performance - GCN~\cite{yasaei2022golden} achieves 98.9/88.4\% with localization support, while Topology-GNN~\cite{hassan2023circuit} and basic GNN~\cite{yasaei2021gnn4tj} achieve 93.35/91.38\% and 92/97\% respectively without localization capabilities. Language-based methods and recent gradient boosting approaches demonstrate high effectiveness, with NLP-based methods~\cite{li2024pinpointing} reaching 97.94/97.77\% and decision-tree gradient boosting~\cite{fan2024efficient} achieving 99/98\%, both supporting localization. Graph-theoretical approaches show mixed results - network analysis~\cite{islam2019socio} achieves 98/98\% with localization support, while graph matching~\cite{piccolboni2017efficient} and graph isomorphism~\cite{demrozi2017exploiting} report 100\% recall but lack localization capabilities. Traditional machine learning methods like gradient boosting~\cite{han2019hardware} and ML-Immune~\cite{zareen2018detecting} show varying performance (100\% and 87/85\% respectively), while flow analysis~\cite{nahiyan2017hardware} achieves perfect recall but with significantly longer processing time (292.85s). While our LLM approach has a relatively higher processing time of 10.91s compared to some newer techniques, it offers a better balance of detection accuracy and localization capability.

\subsection{Limitations}
Most of the limitations of this work have to do with those imposed by using LLMs. These issues include token count impacting scalability, which limits the handling of larger designs (a possible workaround could be partitioning larger designs). The probabilistic behavior of LLMs imposes non-determinism, HT signatures leaking out, and potential circumvention. The perturbation engine is implementing basic code transformations. A small sample of benchmark RTL designs to test our ideas on. The cost for running locally vs. the risks of using the cloud and risk privacy for critical designs.  


\section{Conclusion and Future Work}
\label{sec:conclusion}
This work represents one of the first systematic investigations into leveraging LLMs for HT detection in RTL designs. Our experimental results reveal three critical findings. First, Gemini 1.5's perfect precision/recall (100\%/100\%), even under obfuscation, demonstrates LLMs' ability to detect HTs beyond simple historical HT design memorization. Second, we identified a detection-localization gap: models maintain good precision/recall in detecting HT presence (Gemini 1.5 pro: 100\%/100\%, GPT-4o: 100\%/85.7\%, Llama 3.1: 66.7\%/85.7\%) but struggle with precise localization under obfuscation (TLC: 0.72-0.86, PLC: 0.20-0.33). Finally, the consistently high detection rates (especially of Gemini 1.5 pro and GPT-4o) validate LLM approaches as a promising complement to existing HT detection methods while highlighting areas for improvement in localization precision and HT categorization.

Future work will focus on increasing the dataset to include more designs and either better obfuscate or include designs outside the scope of a trained LLM. This will help clarify whether the agents themselves provide the measured capabilities as emergent skills or the transformer focuses on variations from statistical HDL designs.   






\end{document}